\documentclass[aps,prc,reprint,amsmath]{revtex4-1}

\usepackage{amsfonts}

\renewcommand\bar\overline
\newcommand\gO{\text{O}}
\newcommand\SO{\text{SO}}
\newcommand\lo{\mathfrak{o}}
\newcommand\Cl{\text{Cl}}
\newcommand\tr{\text{tr}\,}
\renewcommand\log{\text{log}\,}
\newcommand\pf{\text{pf}}
\newcommand\sgn{\text{sgn}}

\begin{document}

\title{Fock space representations of Bogolyubov transformations as
  spin representations}

\author{K. Neerg\aa rd}

\affiliation{Fjordtoften 17, 4700 N\ae stved, Denmark}

\begin{abstract}
  The representation on a Fock space of the group of Bogolyubov
  transformations is recognized as the spin representation of an
  orthogonal group. Derivations based on this observation of some
  known formulas for the overlap amplitude of two Bogolyubov
  quasi-fermion vacuum states that are in some cases more complete
  than those in the literature are shown. It is pointed out that the
  name of an ``Onishi formula'' is assigned in the literature to two
  different expressions which are related but have different scopes.
  One of them has what has been described as a sign problem, the other
  one, due to Onishi and Yoshida, has a more limited scope and no sign
  problem. I give a short proof of the latter, whose derivation is
  missing in the paper by Onishi and Yoshida, and a new, combinatoric,
  proof of an equivalent formula recently derived by Robledo.
\end{abstract}

\maketitle

\section{\label{sec:intr}Introduction}

Bogolyubov quasi-nucleon vacua and their overlap amplitudes are
ubiquitous in contemporary calculations of nuclear structure. For a
review, see \cite{ref:Rob19}. Methods such as the generator coordinate
method \cite{ref:Hil53} and, more specifically, the projection of a
quasi-nucleon vacuum onto the state space of some conserved quantum
numbers \cite{ref:Pei57} require the calculation of overlap amplitudes
between different vacua. Some formulas for such overlap amplitudes
involve a square root. This gives rise to a sign ambiguity that has
been called in the literature ``the sign problem of the Onishi
formula'' or the like \cite{ref:Sch04,ref:Oi05,ref:Ben09,ref:Rob09,
  ref:Ave12,ref:Ber12,ref:Miz12,ref:Miz18,ref:Bal18,ref:Por22},
referring to a paper by Onishi and Yoshida \cite{ref:Oni66}. This sign
ambiguity was first discussed by W\"ust and me after being recognized
in the context of quantum number projection of cranked quasi-nucleon
vacua \cite{ref:Nee83}. (Our paper has an unfortunate error. From
(3.4) onwards, every matrix transposition except the second one in the
first line on page 321 should be a Hermitian conjugation.) The
ambiguity is shown in the present paper to be related to the well
known double-valuedness of the representation of spatial rotations on
the spin states of a spin $1/2$ fermion, which is the simplest in a
family of double-valued, so-called spin representations of orthogonal
transformations.

A Bogolyubov quasi-nucleon vacuum, or, more generally, quasi-fermion
vacuum, is determined by the Bogolyubov transformation that relates
its annihilation operators to those of the physical vacuum or some
other reference vacuum. In Sec. \ref{sec:rep}, I first recall the
known fact that the group of Bogolyubov transformations related to
some finite-dimensional space of single-fermion states is isomorphic
to an orthogonal group, and then briefly review the theory of spin
representations. The relation to physics is established by the
observation that the Fock space, that is, the space of states of any
number of fermions inhabiting the space of single-fermion states, has
the structure of a spinor space and therefore carries a spin
representation of the group. In most applications, Bogolyubov
transformations are assumed to be unitary with unit determinant. At
the end of Sec. \ref{sec:rep}, I give the details of a derivation of
an expression for the spin-representation image of such a Bogolyubov
transformation that appears in \cite{ref:Nee83} without a proof.

This expression forms the basis for a derivation in Sec. \ref{sec:vac}
of a formula for the overlap amplitude of two arbitrary members of a
large class of quasi-fermion vacua in terms of their generating
Bogolyubov transformations what is called the ``Onishi formula'' in
the much cited book by Ring and Schuck \cite{ref:Rin80}. My derivation
does not require certain restricting assumptions made there and in
some earlier work. I also rederive some formulas in the literature for
the matrix element between two Bogolyubov vacua of the transformation
of the Fock space generated by a unitary transformation of the
single-fermion state space, and show that all these formulas have an
unavoidable sign ambiguity due to the double-valuedness of the spin
representation.

The ``Onishi formula'' of Ring and Schuck is not the formula of Onishi
and Yoshida. Their formula for the overlap amplitude of two
quasi-fermion vacua has a more limited scope as it applies only when
these vacua have non-zero overlaps with the reference vacuum. Due to
this restriction, it has no sign ambiguity. Notably, Onishi and
Yoshida do not show a derivation of their formula. I give in
Sec.~\ref{sec:oni} a short proof of it by repeated application of a
relation  mentioned in their paper.

W\"ust and I devise in \cite{ref:Nee83} a method to calculate
numerically the unambiguous overlap amplitude of Onishi and Yoshida.
Recently, Robledo proposed another method which may be numerically
more stable \cite{ref:Rob09}. I show in Sec.~\ref{sec:rob} that
Robledo's formula can be derived directly from that of Onishi and
Yoshida. Robledo's derivation is based on Berezin integration, and
recently, some other derivations appeared \cite{ref:Miz18,ref:Por22}.
Before summarizing the paper in Sec.~\ref{sec:sum}, I show at the end
of Sec.~\ref{sec:rob} yet another, combinatoric, derivation.

\section{\label{sec:rep}Fock space representation of the Bogolyubov
  group}

Most structure calculations in atomic, molecular and nuclear physics
employ a space $\mathcal S$ of single-fermion states of finite
dimension $d$. Corresponding to orthonormal basic states $| i \rangle$
in $\mathcal S$ one can define annihilation operators $a_i$. I call
\textit{field operators} the linear combinations $\alpha$ of $a_i$ and
$a^\dagger_i$ and denote by $\mathcal F$ the $2d$-dimensional space of
these operators. A \textit{Bogolyubov transformation} \cite{ref:Bog58}
is a linear transformation of $\mathcal F$ that preserves the
anticommutator $\{\alpha,\beta\}$. Since the anticommutator is a
symmetric bilinear form, the group of Bogolyubov transformations,
which I call the \textit{Bogolyubov group}, is isomorphic to the group
$\gO(2d)$ of orthogonal transformations in $2d$ complex dimensions,
and I identify the Bogolyubov group with $\gO(2d)$. Given any basis
for $\mathcal F$, I denote by $\alpha_-$ and $\alpha_|$ the row and
column of the basic field operators. The \textit{coordinate
  representation} $g \mapsto G$ of $\gO(2d)$ in this basis is then
defined by $g \alpha_- = \alpha_- G$, so that
$g_1 g_2 \mapsto G_1 G_2$, where $G$ is a $2d \times 2d$ matrix. This
is a faithful representation; that is, $G$ determines $g$.

Usually a Bogolyubov transformation is assumed to be also
\textit{unitary} in the sense that it preserves the Hermitian inner
product $\{\alpha^\dagger,\beta\}$ in $\mathcal F$. See, however,
\cite{ref:Bal69} for an exception. The Hermitian field operators
$\alpha_{i+} = a_i + a^\dagger_i$ and
$\alpha_{i-} = -i (a_i - a^\dagger_i)$ obey
$\{\alpha_{is},\alpha_{i's'}\} = \{\alpha^\dagger_{is},\alpha_{i's'}\}
= 2 \delta_{is,i's'}$. (The use of $i$ both as an index and to denote
the imaginary unit should cause no confusion.) In this basis, the
condition of simultaneous orthogonality and unitarity reads
$G^T G = G^\dagger G = 1$, where 1 denotes the unit matrix. Hence $G$
is real. This implies, in particular, that
$g \alpha^\dagger = (g \alpha)^\dagger$. The group of unitary
Bogolyubov transformations, the \emph{unitary Bogolyubov group}, is
thus isomorphic to the group $\gO(2d,\mathbb R)$ of orthogonal
transformations in $2d$ real dimensions associated with a positive
definite quadratic form, and I identify the unitary Bogolyubov group
with $\gO(2d,\mathbb R)$. This isomorphism of the unitary subgroup of
a complex orthogonal group to the real subgroup was pointed out by
Weyl \cite{ref:Wey39}, and the isomorphism of the general and unitary
Bogolyubov groups to the complex and real orthogonal groups in $2d$
dimensions was noticed by Balian and Brezin \cite{ref:Bal69}. Most of
the discussion of $\gO(2d)$ in the present section applies almost
verbally also to $\gO(2d,\mathbb R)$. I shall in general not mention
explicitly the modifications pertaining to this case.

The group $\gO(2d)$ has a maximal connected subgroup $\SO(2d)$ of
index 2. The transformations in $\SO(2d)$ have determinant 1 and those
in its, also connected, coset determinant $-1$ \cite{ref:Wey39}. That
these sets are connected means that there is a continuous path between
any two elements. The elements of $\SO(2d)$ may be seen as rotations
of the $2d$-dimensional space and those of the coset as combinations
of a rotation and a reflection. These two types of orthogonal
transformations are sometimes called proper and improper, and one can
distinguish accordingly between proper and improper Bogolyubov
transformations.

The \textit{Fock space} $\mathcal K$ associated with $\mathcal S$ is
the space of states formed by the action of a polynomial in the
creation operators $a_i^\dagger$ on a state killed by the annihilation
operators $a_i$ and conceived as a vacuum state. It has the structure
of a \textit{spinor space} \cite{ref:Bra35}. The operators on
$\mathcal K$ form an algebra isomorphic to the Clifford algebra
\cite{ref:Cli78} $\Cl(2d)$, and I identify the algebra of operators on
$\mathcal K$ with $\Cl(2d)$. Every element of $\Cl(2d)$ can then be
written as a a polynomial in field operators. It was shown by Brauer
and Weyl that $\gO(2d)$ has a \emph{double-valued} representation on
the spinor space $\mathcal K$, the so-called \textit{spin
  representation} \cite{ref:Bra35}. Specifically, this representation
maps every $g \in \gO(2d)$ to a \emph{pair} $\pm \bar g$ of elements
of $\Cl(2d)$ in such a way that
$g_1 g_2 \mapsto \pm \bar g_1 \bar g_2$, and there is a path in
$\SO(2d)$ from the identity 1 back to itself which connects its images
$\pm 1$ continuously. As pointed out by Weyl \cite{ref:Wey39}, the
second property implies that the connected sets, $\SO(2d)$ and its
improper coset, are not \textit{simply} connected: this path cannot be
contracted to a point. That (unitary) Bogolyubov groups are not simply
connected was noticed in \cite{ref:Nee83}.

The representation $g \mapsto \pm \bar g$ is defined in
\cite{ref:Bra35} by
\begin{equation}\label{eq:(ga)barg=barga}
  (g \alpha) \bar g = \bar g \alpha
\end{equation}
and
\begin{equation}\label{eq:(taug)g=1}
  (\tau \bar g) \bar g = 1 ,
\end{equation}
where $\tau$ is the linear operator on $\Cl(2d)$ that inverts the
order of the factors in every product of field operators. For a given
$g$, the condition \eqref{eq:(ga)barg=barga} determines $\bar g$
within a numeric factor, and the condition \eqref{eq:(taug)g=1} fixes
this factor up to a sign. Since both conditions are compatible with
the group relations, they thus define a \emph{possibly} double-valued
\emph{representation}. The fact that this representation turns out
\emph{actually} double-valued implies that \emph{the entire Bogolyubov
  group (or its unitary subgroup) cannot be mapped continuously and
  single-valuedly into the space of linear transformations of the Fock
  space in such a way that this map $g \mapsto \bar g$ obeys
  \eqref{eq:(ga)barg=barga} for every $g$ and $\alpha$. This will turn
  out in Sec.~\ref{sec:vac} to be the origin of the so-called sign
  problem of the Onishi formula mentioned in the introduction.} In
fact, \eqref{eq:(ga)barg=barga} alone allows the normalized $\bar g$
to be multiplied by a $g$-dependent numeric factor, but for the
resulting map to be continuous, this factor must depend continuously
on $g$, so the resulting map remains double-valued. From now on,
$\bar g$ is understood to satisfy both conditions
\eqref{eq:(ga)barg=barga} and \eqref{eq:(taug)g=1}.

The space $\Cl(2d)$ of operators on $\mathcal K$ is the direct sum of
subspaces $\Cl_\pm(2d)$ formed by the operators that can be expressed
by polynomials in field operators whose terms have only even and only
odd degree, respectively. Evidently
$\Cl_\pm(2d) \Cl_\pm(2d) = \Cl_+(2d)$ and
$\Cl_+(2d) \Cl_-(2d) = \Cl_-(2d) \Cl_+(2d) = \Cl_-(2d)$. For the
reflection $g$ of $\mathcal F$ along the direction of the vector
$\alpha_{1+}$, that is,
\begin{equation}
  \quad \alpha_{1+} \mapsto - \alpha_{1+} , \quad
  \alpha_{is} \mapsto \alpha_{is} \text{ for } is \ne 1+ ,
\end{equation}
the operator
\begin{equation}\label{eq:refl}
  \bar g = \prod_{is \ne 1+} \alpha_{is} ,
\end{equation}
obeys \eqref{eq:(ga)barg=barga} and \eqref{eq:(taug)g=1}. The order of
the factors in \eqref{eq:refl} is immaterial since reordering changes
at most the sign of the product. Orthogonal transformation gives an
analogous expression for a spin-representation image of any other
reflection, and all these images belong to $\Cl_-(2d)$. Since every
proper orthogonal transformation is the product of an even number of
reflections, and every improper orthogonal transformation a product of
an odd number of reflections \cite{ref:Car38}, it follows that the
spin representation maps $\SO(2d)$ into $\Cl_+(2d)$, and its improper
coset into $\Cl_-(2d)$. This construction succeeds in
$\gO(2d,\mathbb R)$. There, the spin-representation image of every
reflection is unitary, whence it follows that the image of every $g$
is unitary. The spin representation of $\gO(2d)$ is irreducible but
splits upon restriction to $\SO(2d)$ into two inequivalent irreducible
components carried by the subspaces $\mathcal K_\pm$ of $\mathcal K$
with even and odd particle number, respectively \cite{ref:Bra35}.

The \textit{Lie algebra} of a continuous linear group is the
anticommutative algebra (the \emph{definition} of a Lie algebra) of
infinitesimal deviations of group elements from the identity 1 with
the commutator product \cite{ref:Wey39}. I denote by $\lo(2d)$ the Lie
algebra of $\gO(2d)$ and identify it with its realization on
$\mathcal F$. The Lie algebra $\lo(2d)$ then consists of the linear
transformations $\alpha \mapsto x \alpha$ of $\mathcal F$ that obey
$\{x \alpha , \beta\} + \{\alpha , x \beta\} = 0$. In the basis of
field operators $\alpha_{is}$, the matrices $X$ of its coordinate
representation $x \mapsto X$ are skew symmetric. Upon restriction to
$\lo(2d,\mathbb R)$ they are also real. In the basis
$a_1,\dots,a_d,a_1^\dagger,\dots,a_d^\dagger$, they obey
\begin{equation}\label{eq:basisaa+}
  X \begin{pmatrix}0&1\\1&0\end{pmatrix}
  = - \begin{pmatrix}0&1\\1&0\end{pmatrix} X^T ,
\end{equation}
in terms of block matrices with $d \times d$ blocks, and $X$ is
anti-Hermitian upon restriction to $\lo(2d,\mathbb R)$. The spin
representation of $\gO(2d)$ gives rise in the infinitesimal limit to
two inequivalent irreducible (single-valued) spin representations of
$\lo(2d)$ carried by the same subspaces of $\mathcal K$ as those of
$\SO(2d)$ \cite{ref:Bra35}. These representations were discovered in
their abstract forms by Cartan before the work of Brauer and Weyl
\cite{ref:Car13}.

Let $x$ denote an element of $\lo(2d)$, let $X$ be the corresponding
matrix in the coordinate representation pertaining to some basis for
$\mathcal F$, and let $\bar x$ be the spin representation image of
$x$. Inserting $g = 1 + x$ in \eqref{eq:(ga)barg=barga} and
linearizing in $x$ gives
\begin{equation}\label{eq:defxbar}
  x \alpha_- = \alpha_- X = [ \bar x , \alpha_- ] . 
\end{equation}
Since in a neighborhood of 1, the transformation $g$ belongs to
$\SO(2d)$, the operator $\bar x$ belongs to $\Cl_+(2d)$, and in order
that $[ \bar x , \alpha_- ]$ be linear in the field operators, it must
then be given by a quadratic polynomial in field operators. In the basis
$a_1,\dots,a_d,a_1^\dagger,\dots,a_d^\dagger$, the complete solution
of the last equation in \eqref{eq:defxbar} is then
\begin{equation}
  \bar x = \tfrac12
    \alpha_- X \begin{pmatrix}0&1\\1&0\end{pmatrix} \alpha_| + \gamma
\end{equation}
with an arbitrary numeric constant $\gamma$. The normalization
\eqref{eq:(taug)g=1} gives $\gamma = 0$, so
\begin{equation}\label{eq:barx}
  \bar x = \tfrac12
    \alpha_- X \begin{pmatrix}0&1\\1&0\end{pmatrix} \alpha_| .
\end{equation}
Note that the matrix sandwiched here between $\alpha_-$ and
$ \alpha_|$ is skew symmetric due to \eqref{eq:basisaa+}. Given the
representation $x \mapsto \bar x$, one can determine $\bar g$ for
$g \in \SO(2d)$ by integrating the differential equation
$\bar g'(t) = \bar x(t) \bar g(t)$ along a path
$g(t), \, 0 \le t \le 1,$ such that $g(0) = 1$, $g(1) = g$, and
$g'(t) = x(t) g(t)$. Choosing different paths gives the two solutions
for $\bar g$ with opposite signs.

Now assume $g \in \SO(2d,\mathbb R)$. I the basis of field operators
$\alpha_{is}$, the matrix $G$ is then real orthogonal, and $|G| = 1$,
so $G$ is real-orthogonal equivalent to a block diagonal matrix with
diagonal blocks of the form
\begin{equation}\label{eq:2x2}
  \begin{pmatrix}\cos\phi&\sin\phi\\-\sin\phi&\cos\phi\end{pmatrix}
  = \exp \begin{pmatrix}0&\phi\\-\phi&0\end{pmatrix} ,
\end{equation}
where $\phi$ is real. It follows that $G = \exp X$, where $X$ is real
and skew symmetric and thus represents an element $x$ of
$\lo(2d,\mathbb R)$. The relation $G = \exp X$ translates to
$g = \exp x$. Because $g(t) = \exp t x, \, 0 \le t \le 1,$ defines a
path from 1 to $g$ with $g'(t) = x g(t)$, this translates, in turn, to
$\bar g = \exp \bar x$, where $\bar g$ is one of the two
spin-representation images of $g$. In the basis
$a_1,\dots,a_d,a_1^\dagger,\dots,a_d^\dagger$, the expression
\eqref{eq:barx} gives
\begin{equation}\label{eq:barg}
  \bar g = \exp \tfrac12 
    \alpha_- X \begin{pmatrix}0&1\\1&0\end{pmatrix} \alpha_| .
\end{equation}
This is essentially (2.11) in \cite{ref:Nee83}, where $X$ is written
as $\log G$. The angles $\phi$ in \eqref{eq:2x2} are not unique; the
same $G$ results when an arbitrary integral multiple of $2 \pi$ is
added to each $\phi$. In \eqref{eq:barg}, this gives rise, exactly, to
the sign ambiguity of $\bar g$. In fact, when just one of the angles
$\phi$ varies continuously from 0 to $2 \pi$ while the rest are kept
at 0, the transformation $g$ goes from 1 back to itself, but $\bar g$
goes from 1 to $-1$. Since this path runs within $\gO(2d,\mathbb R)$,
the double-valuedness persists upon restriction to this subgroup
\cite{ref:Bra35}.

Adding an integral multiple of $2 \pi$ to each $\phi$ corresponds in
the basis $a_1,\dots,a_d,a_1^\dagger,\dots,a_d^\dagger$ to adding an
arbitrary integral multiple of $2 \pi i$ to each eigenvalue of the
anti-Hermitian matrix $X$. Bally and Duguet propose to choose for
these eigenvalues always the principal logarithms of those of $G$
\cite{ref:Bal18}. This renders $\bar g$ discontinuous at the branch
cuts of the logarithms, and the map $g \mapsto \bar g$ will not
preserve the group relations.

\section{\label{sec:vac}Quasi-fermion vacua and their overlaps}

In the structure calculations, one is specifically interested in the
\textit{quasi-fermion vacuum states} $| g \rangle$, which are states
annihilated by the transformed annihilation operators $g a_i$. It
follows from \eqref{eq:(ga)barg=barga} that
$| g \rangle \propto \bar g | \rangle$, where $| \rangle$ denotes a
``reference'' state annihilated by the field operators $a_i$. The
latter need not be thought of as representing the physical vacuum. As
pointed out by Bally and Duguet \cite{ref:Bal18}, every state
proportional to some $\bar g | \rangle$ may serve as a reference
state. This follows from the Bogolyobov transformations forming a
group. When $| \rangle$ is fixed, the double-valuedness of the map
$g \mapsto \bar g$ implies double-valuedness of the map
$g \mapsto \bar g | \rangle$. An argument as in Sec. \ref{sec:rep}
shows that the double-valuedness persists if $\bar g | \rangle$ is
multiplied by a $g$-dependent numeric factor, as done, for example, in
the theory of Bally and Duguet. Therefore \emph{the entire Bogolyubov
  group (or its unitary subgroup) cannot be mapped continuously and
  single-valuedly into the Fock space in such a way that this map
  $g \mapsto | g \rangle$ obeys $(g a_i) | g \rangle = 0$ for every
  $g$ and $i$.} Since any $g$-dependent numeric factor is well-defined
in a given formalism, it is interesting to analyze the bare expression
$| g \rangle = \pm \bar g | \rangle$, which implies
$| g_1 g_2 \rangle = \pm \bar g_1 | g_2 \rangle$. So from now on, this
is the definition of $| g \rangle$.

As noted in the introduction, overlap amplitudes
$\langle g_1 | g_2\rangle$ are central in many types of calculations.
I limit my discussion of such amplitudes to \emph{unitary} Bogolyubov
transformations. Then every $\bar g$ is unitary, whence follows
$\bar {g^{-1}} = \bar g^{-1} = \bar g^\dagger$. I also assume
$\langle | \rangle = 1$, which then implies
$\langle g | g \rangle = 1$ for every $g$. The relation
$\langle g_1 | g_2\rangle = \langle | \bar g_1^\dagger \bar g_2 |
\rangle = \langle | \bar g_1^{-1} \bar g_2 | \rangle = \langle | \bar
{ g_1^{-1} g_2 } | \rangle$ reduces the calculation of
$\langle g_1 | g_2\rangle$ to the calculation of some
$\langle | \bar g | \rangle$ \cite{ref:Rin80}. One can assume
$g \in \SO(2d,\mathbb R)$ because otherwise $\bar g \in \Cl_-(2d)$,
whose elements connect the subspaces $\mathcal K_\pm$ of $\mathcal K$,
whence $\langle | \bar g | \rangle = 0$ because
$| \rangle \in \mathcal K_+$. When also \emph{$d$ is even}, as usual
in nuclear structure calculations, a formula for
$\langle | \bar g | \rangle$ can then be derived from the
Bloch-Messiah decomposition \cite{ref:Blo62} of the matrix $G$
representing $g$ in the coordinate representation pertaining to the
basis $a_1,\dots,a_d,a_1^\dagger,\dots,a_d^\dagger$. In the notation
of Beck, Mang, and Ring \cite{ref:Bec70}, this decomposition reads
\begin{equation}\label{eq:Bloch}
  G = \begin{pmatrix} D^* & 0 \\ 0 & D \end{pmatrix}
    \begin{pmatrix} U & V \\ V & U \end{pmatrix}
    \begin{pmatrix} C^* & 0 \\ 0 & C \end{pmatrix} ,
\end{equation}
where $D$ and $C$ are unitary and $U$ and $V$ are block diagonal with
$2 \times 2$ diagonal blocks
\begin{equation}
  \begin{pmatrix} u & 0 \\ 0 & u \end{pmatrix}
   \quad \text{and} \quad
   \begin{pmatrix} 0 & v \\ -v & 0 \end{pmatrix} .
\end{equation}
Here, $u$ and $v$ are non-negative and obey $u^2 + v^2 = 1$. Each of
the three factors $\Gamma$ in \eqref{eq:Bloch} is unitary and has unit
determinant. Further, each of them obeys the condition of
orthogonality
\begin{equation}
  \Gamma \begin{pmatrix}0&1\\1&0\end{pmatrix} \Gamma^T
  = \begin{pmatrix}0&1\\1&0\end{pmatrix} ,
\end{equation}
which for a unitary $\Gamma$ is equivalent to 
\begin{equation}
  \Gamma \begin{pmatrix}0&1\\1&0\end{pmatrix}
  = \begin{pmatrix}0&1\\1&0\end{pmatrix} \Gamma^* .
\end{equation}
Thus each of them represents a proper unitary Bogolyubov
transformation, which I denote by $g_D$, $g_W$, and $g_C$,
respectively. I set out to analyze the action of each of these three
factors of the product $g = g_D g_W g_C$.

To calculate the action of $g_C$, let the basic single-fermion states
$| i \rangle$ be chosen such that $C$ is diagonal. Since $C$ is
unitary, one can set $C = \exp Y$, where $Y$ is diagonal and
imaginary. I denote by $y_i$ the diagonal entries in $Y$. The
rightmost matrix in \eqref{eq:Bloch} becomes $\exp X$, where $X$ is
block diagonal with $-Y$ in the upper diagonal block and $Y$ in the
lower diagonal block. Then \eqref{eq:barx} gives
\begin{equation}\label{eq:barxc}
  \bar x = \tfrac12 \sum_i y_i
    \left( a_i^\dagger a_i - a_i a_i^\dagger  \right)
  = \sum_i y_i \left( a_i^\dagger a_i - \tfrac12  \right) ,
\end{equation}
whence
\begin{equation}
  \bar g_C \, | \rangle = \exp \bar x \, | \rangle
  = \exp \left( -\tfrac12 \sum y_i \right) | \rangle
  = \sqrt{|C^*|} \, | \rangle .
\end{equation}
Analogously,
$\langle| \, \bar g_D = \sqrt{|D^*|} \, \langle|$.

One can set
\begin{equation}
  \begin{pmatrix} U & V \\ V & U \end{pmatrix} = \exp X ,
\end{equation}
where
\begin{equation}
  X =  \begin{pmatrix} 0 & Y \\  Y & 0 \end{pmatrix} ,
\end{equation}
and $Y$ is block diagonal with $2 \times 2$ diagonal blocks
\begin{equation}
  \begin{pmatrix} 0 & y \\ - y & 0 \end{pmatrix}
\end{equation} 
such that $\cos y = u$ and $\sin y = v$. With $y_i$ denoting the entry
of $Y$ with indices $i,i+1$, where $i$ is odd, one gets from
\eqref{eq:barx} that
\begin{equation}\label{eq:logw}
  \bar x= \sum_\text{odd $i$} y_i
    \left( a_i a_{i+1} + a_i^\dagger a_{i+1}^\dagger \right) .
\end{equation}
By 
\begin{multline}
  \left( a_i a_{i+1} + a_i^\dagger a_{i+1}^\dagger \right)^2
    | \rangle \\
  = \left( a_i a_{i+1} a_i^\dagger a_{i+1}^\dagger
      + a_i^\dagger a_{i+1}^\dagger a_i a_{i+1} \right) | \rangle
  = - | \rangle
\end{multline}
follows
\begin{equation}\label{eq:<w>}
  \langle | \bar g_W | \rangle = \langle | \exp \bar x | \rangle
  = \prod_\text{odd $i$} \cos y_i
  = \prod_\text{odd $i$} u_i = \sqrt{|U|} ,
\end{equation}
where $u_i$ are the diagonal entries of $U$, and the non-negative
square root is taken in the last expression.

Combining these results gives
\begin{equation}\label{eq:Ring}
  \langle | \bar g | \rangle = \sqrt{|D^* U C^*|} = \sqrt{|A^*|}
\end{equation}
in the notation of \cite{ref:Bec70}, where
\begin{equation}\label{eq:ABBA}
  G = \begin{pmatrix} A^* & B \\ B^* & A \end{pmatrix} .
\end{equation}
The appearance of a square root in \eqref{eq:Ring} reflects the sign
ambiguity of $\bar g$ . In the derivation above, it stems from the
multi-valuedness of $Y$ as a solution of $C = \exp Y$ or $D = \exp Y$.
In \cite{ref:Rin80}, the identity $\langle g | \rangle = \sqrt{|A|}$,
which is equivalent to \eqref{eq:Ring}, is derived in the case
$D = C = 1$, which may be generalized by a change of basis for
$\mathcal S$ to the case when $A$ is Hermitian and positive
semi-definite. In that case, $|A^*|= |A| \ge 0$.

From \eqref{eq:Ring} and \eqref{eq:ABBA}, one gets immediately
\begin{multline}\label{eq:Beck}
  \langle g_1 | g_2\rangle
  = \langle | \bar g_1^{-1} \bar g_2 | \rangle \\
  = \sqrt{ | A_1^T A_2^* + B_1^T B_2^* | }
  = \sqrt{ | A_2^\dagger A_1 + B_2^\dagger B_1 | } .
\end{multline}
The last expression is (2.14a) in \cite{ref:Bec70}, where it is
derived from \eqref{eq:oni} with a specific choice of $D_1$, $C_1$,
$D_2$, and $C_2$. In \cite{ref:Rin80}, it is derived from
$\langle g | \rangle = \sqrt{|A|}$.

In quantum number projection, one needs matrix elements
$\langle g_1 | \bar u | g_2 \rangle$, where $\bar u$ is the unitary
transformation of $\mathcal K$ generated by a unitary transformation
$u$ of $\mathcal S$. Explicitly,
$\bar u = \exp a^\dagger_- \, Y \, a_|$, where $u = \exp y$ and
$y | i \rangle_- = | i \rangle_- Y$ with $| i \rangle_-$ denoting the
row of states $| i \rangle$, and $ a^\dagger_-$ is the row of
operators $a_i^\dagger$, and $a_|$ the column of operators $a_i$. The
operator
\begin{equation}
  a^\dagger_- \, Y \, a_|
  = \tfrac12 \left( \alpha_-
    \begin{pmatrix} 0 & - Y^T \\ Y & 0 \end{pmatrix}
    \alpha_| + \tr Y \right)
\end{equation}
with $\alpha_-$ and $\alpha_|$ as in \eqref{eq:barx} is, except for
the last term in the brackets, the spin representation image $\bar x$
of the element $x$ of $\lo(2d,\mathbb R)$ with
\begin{equation}
  X = \begin{pmatrix} - Y^T & 0 \\ 0 & Y \end{pmatrix}
    = \begin{pmatrix} Y^* & 0 \\ 0 & Y \end{pmatrix} ,
\end{equation}
where the last transformation stems from $X$ being anti-Hermitian.
Therefore $\bar u \exp - \tfrac12 \tr Y = \bar u / \sqrt{| U |}$, with
$U = \exp Y$, is the spin representation image of the element $g$ of
$\SO(2d,\mathbb R)$ determined by
\begin{equation}
  G = \begin{pmatrix} U^* & 0 \\ 0 & U \end{pmatrix} .
\end{equation}
Using
$| U | = | U^T |$, one arrives at
\begin{multline}
  \langle g_1 | \bar u | g_2 \rangle
  = \langle | \bar g_1^{-1} \bar g \, \bar g_2 | \rangle
    \sqrt{| U^T |} \\
  = \sqrt{ | ( A_1^T U^*  A_2^*
      + B_1^T U B_2^* ) U^T | } .
\end{multline}
This is (2.13) in \cite{ref:Nee83} (with missing equation number),
whence the first expression in \eqref{eq:Beck} is a special case. For
$g_1 = g_2$ it is (34) in \cite{ref:Goe72} except that there, the
factor $\sqrt{| U^T |}$ gets lost in the derivation from the preceding
equation (33).

\section{\label{sec:oni}Formula of Onishi and Yoshida}

Ring and Schuck call $\langle g | \rangle = \sqrt{|A|}$ or the second
expression in \eqref{eq:Beck} the ``Onishi formula'' \cite{ref:Rin80},
referring to the paper \cite{ref:Oni66} by Onishi and Yoshida. As
announced in the introduction, the formula actually written by Onishi
and Yoshida is quite different. First of all, these authors consider,
not the state $| g \rangle$, but the state
\begin{equation}
  | \tilde g \rangle = \frac {| g \rangle} {\langle | g \rangle} .
\end{equation}
This state has no sign ambiguity; the arbitrary sign in $| g \rangle$
cancels out in the division. The state $| \tilde g \rangle$ can be
defined alternatively by $(g a_i)| \tilde g \rangle = 0$ and the
normalization $\langle | \tilde g \rangle = 1$ (which implies that
generally $\langle \tilde g | \tilde g \rangle > 1$). The uniqueness
of $| \tilde g \rangle$ comes at the cost of a lack of generality;
$| \tilde g \rangle$ is undefined when $\langle | g \rangle = 0$. From
another point of view, when $G$ is written as in \eqref{eq:Bloch}, the
maximal neighborhood of 1 where $\langle | g \rangle \ne 0$ is
described by $y_i < \pi/2$ for every $y_i$ in \eqref{eq:<w>}, and this
set is simply connected. Bally and Duguet introduce in their formalism
\cite{ref:Bal18} a similar limitation by demanding that for all
Bogolyubov transformations $g$ to be considered, the amplitudes
$\langle | \text{vac} (g) \rangle$ have equal phases, where
$| \text{vac} (g) \rangle$ is the quasi-fermion vacuum state assigned
to $g$. This condition clearly fails when
$\langle | \text{vac} (g) \rangle = 0$.

Onishi and Yoshida consider the case when every $g$ is unitary and
proper, and express $| \tilde g \rangle$ by its Thouless expansion
\cite{ref:Tho60}
\begin{equation}
  | \tilde g \rangle = \exp \tfrac12 a^\dagger_- F a^\dagger_| \, 
    | \rangle ,
\end{equation}
where the skew symmetric matrix $F$ is related to the matrices $A$ and
$B$ in \eqref{eq:ABBA} by $F = ( B A^{-1} )^*$ \cite{ref:Goe73} (which
shows once more way that $| \tilde g \rangle$ is undefined when
$| A | = 0$). They hence derive
\begin{equation}\label{eq:oni}
  \langle \tilde g_1 | \tilde g _2 \rangle
  = \exp \tfrac12 \tr \log ( 1 + F_1^\dagger F_2 ) .
\end{equation}
(To be accurate, in their formula, the second term in the argument of
the logarithm is $F_2 F_1^\dagger$, but this makes no difference due
to the trace.) Notably, Onishi and Yoshida do not show their
derivation. The short derivation below makes repeated use of the
relation
\begin{equation}
  \big[ a_| , \exp \tfrac12 a^\dagger_- F a^\dagger_| \big]
  =  F a^\dagger_| \exp \tfrac12 a^\dagger_- F a^\dagger_| ,
\end{equation}
which is mentioned in their paper.

What must be proven is that for any two skew symmetric matrices $P$
and $Q$, the identity
\begin{multline}\label{eq:tbp}
  \omega := \Big\langle \Big| \exp \tfrac12 a_- P a_| \,
    \exp \tfrac12 a^\dagger_- Q a^\dagger_| \Big| \Big\rangle \\
  = \exp \tfrac12 \tr \log (1 + P Q) ,   
\end{multline}
holds. With
\begin{equation}\label{eq:f}
  f(z): = \Big\langle \Big| \exp \tfrac12 z a_- P a_| \,
    \exp \tfrac12 \, a^\dagger_- Q a^\dagger_|
    \Big| \Big\rangle ,
\end{equation}
one gets
\begin{equation}\begin{aligned}
  f'(z) 
  & = \tfrac12 \Big\langle \Big| \big( \exp \tfrac12 z a_- P a_| \big)
        a_- P a_| \exp \tfrac12  a^\dagger_- Q a^\dagger_| 
        \Big| \Big\rangle \\
  & = \tfrac12 \Big\langle \Big| \big( \exp \tfrac12 z a_- P a_| \big) 
        a_- P Q a^\dagger_|  \exp \tfrac12  a^\dagger_- Q a^\dagger_|
        \Big| \Big\rangle \\
  & = \tfrac12 \Big\langle \Big| \big( \exp \tfrac12 z a_- P a_| \big)
  \\ & \hskip6em \big( \tr P Q - a^\dagger_- Q P a_| \big)
       \exp \tfrac12  a^\dagger_-  Q a^\dagger_| \Big| \Big\rangle \\
  & = \tfrac12 \Big\langle \Big| \big( \exp \tfrac12 z a_- P a_| \big) 
  \\  & \hskip5em  \big( \tr P Q - z a_- P Q P a_| \big)
       \exp \tfrac12  a^\dagger_- Q a^\dagger_| \Big| \Big\rangle \\
  & = \tfrac12 \Big\langle \Big| \big( \exp \tfrac12 z a_- P a_| \big) 
        \\
  & \hskip4em \big( \tr P Q - z a_- P Q P Q a^\dagger_| \big)
        \exp \tfrac12  a^\dagger_-  Q a^\dagger_| \Big| \Big\rangle \\
  & = \tfrac12 \Big\langle \Big| \big( \exp \tfrac12 z a_- P a_| \big) 
        \\
  & \hskip4em \big( \tr (P Q - z P Q P Q) + z a^\dagger_- Q P Q P a_| \big) \\
  & \hskip14.5em \exp \tfrac12  a^\dagger_-  Q a^\dagger_| \Big| \Big\rangle \\
  & = \cdots \\
  & = \tfrac12  \tr (P Q - z P Q P Q + z^2 P Q P Q P Q - \cdots) \\
  & \hskip7.5em \Big\langle \Big| \exp \tfrac12 z a_- P a_| 
\exp \tfrac12  a^\dagger_- Q a^\dagger_| \Big| \Big\rangle \\
  & = \tfrac12  \tr P Q (1 + z P Q )^{-1} f(z) \\
  &  = \Big( \frac d {dz}
        \tfrac12  \tr \log (1 + z P Q) \Big) f(z) .
\end{aligned}\end{equation}
Since $f(0) = 1$, hence follows
\begin{equation}\label{eq:res}
  f(z) = \exp \tfrac12 \tr \log  (1 + z P Q) ,
\end{equation}
and \eqref{eq:tbp}, in particular, provided one takes $\log 1 = 0$.

This argument requires that the Taylor expansion of $(1 + z P Q)^{-1}$
converges. This holds when $z$ is numerically less than the reciprocal
of every non-zero characteristic root of $P Q$. However, because the
expansion of $\exp \frac12 z a_- P a_|$ on powers $z^n$ terminates
when $2 n > d$, the function $f$ is a polynomial. Therefore, by
analytic continuation, despite its nominal multi-valuedness, the right
hand side of \eqref{eq:res} is single-valued outside the singularities
of the logarithm, which occur when $z$ equals minus the reciprocal of
a characteristic root, and it can be extended to the singularities so
as to be defined for every $z$ as a continuous function of $z$. It
follows by a similar argument that the right hand side of
\eqref{eq:tbp} is well defined as a continuous function of the entries
of $P$ and $Q$.

In terms of the characteristic roots $r_i$ of $PQ$, the expression
\eqref{eq:res} can be written
\begin{equation}
  f(z) = \exp \tfrac12 \sum_i \log (1 + z r_i)
    = \sqrt {\prod_i (1 + z r_i ) } ,
\end{equation}
where one must take $\sqrt 1 = 1$ and for $z \ne 0$ choose the sign of
the square root that makes it continuous in $z$. Since $f$ is a
polynomial, the square root is a polynomial in $z$, so the
characteristic roots have even multiplicities \cite{ref:Nee83}. The
only assumption was that $P$ and $Q$ be skew symmetric, so this
property of the spectrum of characteristic roots must hold, in fact,
for every product of two skew symmetric complex matrices. This can be
shown more directly as follows. Cayley proved that the determinant of
a skew symmetric matrix $S$ can be written as the square of a
polynomial in its entries which he called its Pfaffian, and which is
denoted usually by $\pf\, S$ \cite{ref:Cal49,ref:Cal52}. Hence, when
$P$ and $Q $ are skew symmetric, one has
\begin{multline}\label{eq:pf^2}
  |1 + z P Q|
    = \begin{vmatrix}1 + z P Q & - z P\\ 0 & 1 \end{vmatrix}
    = \begin{vmatrix} z P & 1\\ -1 & Q\end{vmatrix}
      \begin{vmatrix} Q & -1 \\ 1 & 0\end{vmatrix} \\
    = \begin{vmatrix} z P & 1 \\ -1 & Q \end{vmatrix}
    = \left(\pf \begin{pmatrix} z P & 1 \\ -1 & Q \end{pmatrix}
      \right)^2 ,
\end{multline}
which implies that the non-zero characteristic roots of $PQ$ have even
multiplicities. Yet another proof of this result was given by Oi,
Mizusaki, Shimizu, and Sun under some restricting assumptions
\cite{ref:Oi19}. Using it, one can express the right hand side of
\eqref{eq:tbp} as
\begin{equation}\label{eq:NW}
  {\prod}'_i (1 + r_i ) ,
\end{equation}
where one out of every pair of equal characteristic roots of $P Q$,
counted with multiplicity, is included in the product. This reduces
the calculation of $\langle \tilde g_1 | \tilde g_2 \rangle$ to the
determination of the characteristic roots of $F_1^\dagger F_2$
\cite{ref:Nee83}.

In \cite{ref:Goe72,ref:Goe73}, the expression \eqref{eq:oni} is
written
\begin{equation}\label{eq:goe}
  \langle \tilde g_1 | \tilde g _2 \rangle
  =\sqrt { | 1 + F_1^\dagger F_2 | } .
\end{equation}
In part of the literature, \eqref{eq:oni} or variants such as
\eqref{eq:goe}, where the right hand side is well defined by
continuity, but whose scope was seen to be limited to a neighborhood
of 1, are called the ``Onishi formula''
\cite{ref:Bal18,ref:Ave12,ref:Ber12,ref:Miz12,ref:Miz18,ref:Por22}.
Mizusaki, Oi, and Shimizu derived \eqref{eq:goe} from the linked
cluster theorem \cite{ref:Miz18}, and Porro and Duguet obtained
\eqref{eq:oni} by a diagrammatic method \cite{ref:Por22}.

\section{\label{sec:rob}Robledo formula}

The definition of the Pfaffian of a $2d$-dimensional skew symmetric
matrix $S$ with entries $s_{ij}$ can be written
\begin{equation}\label{eq:pf}
  \pf \, S = \sum_\pi
    \sgn \begin{pmatrix} 1 & 2 & \dots & 2d \\
      i_1 & i_2 & \dots & i_{2d} \end{pmatrix}
    \prod_{\text{odd $\nu < 2d$}} s_{i_\nu i_{\nu+1}} ,
\end{equation}
where the sum runs over partitions
$\pi = \{\{i_1,i_2\},\dots,\linebreak\{i_{2d-1},i_{2d}\}\}$ of
$\{1,2,\dots,2d\}$ \cite{ref:Cal49}. Hence,
\begin{equation}\label{eq:P=0}
  \pf \begin{pmatrix} 0 & 1 \\ -1 & Q \end{pmatrix}
  = (-1)^{d(d-1)/2}
\end{equation}
because in this case, the product in \eqref{eq:pf} is different from 0
only when $\pi = \{\{1,d+1\},\{2,d+2\},\dots,\linebreak\{d,2d\}\}$.
Writing the right hand side of \eqref{eq:res} as in \eqref{eq:goe},
one gets from \eqref{eq:pf^2} that
\begin{equation}
  f(z) = \sqrt { | 1 + z P Q | } = (-)^{d(d-1)/2} \,
    \pf \begin{pmatrix} z P & 1 \\ -1 & Q \end{pmatrix}
\end{equation}
with the correct sign due to \eqref{eq:P=0}. For $z = 1$, this becomes
\begin{equation}\label{eq:rob}
  \omega = (-)^{d(d-1)/2} \,
    \pf \begin{pmatrix} P & 1 \\ -1 & Q \end{pmatrix} ,
\end{equation}
where $\omega$ is defined in \eqref{eq:tbp}. In a slightly different
but equivalent form, this identity was proven by Robledo
\cite{ref:Rob09} by means of Berezin integration \cite{ref:Ber66}.
Pfaffians can be calculated by Householder transformation
\cite{ref:Hou58}, which may be numerically more stable than
determining the characteristic roots of an arbitrary complex matrix
\cite{ref:Rob09}. The matrix in \eqref{eq:rob} is seen to have twice
the dimension of that of $PQ$. Other derivations of \eqref{eq:rob} are
given by Mizusaki, Oi, and Shimizu \cite{ref:Miz18}, and Porro and
Duguet \cite{ref:Por22}. I give yet another, combinatoric, proof.

Denoting the entries of $P$ and $Q$ by $p_{ij}$ and $q_{ij}$, one can
write
\begin{equation}\begin{aligned}
  \exp \tfrac12 a_- P a_| =
   \sum_{\pi_1} &\ \sgn \begin{pmatrix}
      k_1&k_2&\dots&k_{2m_1}\\i_1&i_2&\dots&i_{2m_1}\end{pmatrix} \\
    & \prod_{\text{odd $\nu < 2m_1$}} p_{i_\nu i_{\nu+1}}
    a_{i_\nu} a_{i_\nu+1} , \\
  \exp \tfrac12 a^\dagger_- Q a^\dagger_| =
    \sum_{\pi_2} &\ \sgn \begin{pmatrix}
      l_1&l_2&\dots&l_{2m_2}\\j_1&j_2&\dots&j_{2m_2}\end{pmatrix} \\
    & \prod_{\text{odd $\nu < 2m_2$}} q_{j_\nu j_{\nu+1}}
    a^\dagger_{j_\nu} a^\dagger_{j_\nu+1} ,
\end{aligned}\end{equation}
where the sums run over partitions
$\pi_1 = \{\{i_1,i_2\},\dots,\linebreak\{i_{2m_1-1},i_{2m_1}\}\}$ and
\mbox{
  $\pi_2 = \{\{j_1,j_2\},\dots,\linebreak\{j_{2m_2-1},j_{2m_2}\}\}$}
of even subsets $\{k_1,k_2,\dots,k_{2m_1}\}$ and
$\{l_1,l_2,\dots,l_{2m_2}\}$ of $\{1,2,\dots,d\}$, and
$k_1<k_2<\cdots<k_{2m_1}$ and $l_1<l_2<\cdots<l_{2m_2}$. Inserting
these expressions into the definition of $\omega$ in \eqref{eq:tbp}
gives
\begin{multline}\label{eq:omega}
  \omega = \sum_{\pi_1 \sim \pi_2} \sgn \begin{pmatrix}
      i_1&i_2&\dots&i_{2m}\\j_1&j_2&\dots&j_{2m}\end{pmatrix} \\
    \prod_{\text{odd $\nu < 2m$}} p_{i_\nu i_{\nu+1}} q_{j_{\nu+1 j_\nu}} ,
\end{multline}
where $\pi_1 \sim \pi_2$ is shorthand for
$\{i_1,i_2,\dots,i_{2m_1}\} = \{j_1,j_2,\dots,j_{2m_2}\}$, which
implies $m_1 = m_2 := m$. Next notice that $(-)^{d(d-1)/2}$ is the
signature of the simultaneous permutation of the rows and the columns
that transforms the matrix in \eqref{eq:rob} into the $d \times d$
block matrix $S$ with diagonal blocks
\begin{equation}\label{eq:diag}
   \begin{pmatrix} 0 & 1 \\ -1 & 0 \end{pmatrix}
\end{equation}
and off-diagonal blocks
\begin{equation}\label{eq:off}
   \begin{pmatrix} p_{ij} & 0 \\ 0 & q_{ij} \end{pmatrix} , 
\end{equation}
so by \eqref{eq:pf}, the identity \eqref{eq:rob} can be written
\begin{equation}\label{eq:S}
   \omega = \pf \, S ,
\end{equation}
I set out to calculate $\pf \, S$.

The pairs $\{i_\nu,i_{\nu+1}\}$ in \eqref{eq:pf} can be so chosen that
always $i_\nu < i_{\nu+1}$. Then, when a factor $s_{i_\nu i_{\nu+1}}$
stems from a submatrix \eqref{eq:diag}, it equals 1, so one can omit
these factors from the product and the corresponding columns from the
permutation symbol, which does not alter the signature. When
$s_{i_\nu i_{\nu+1}}$ is $p_{ij}$, then $i_\nu$ and $i_{\nu+1}$ are
odd, and when $s_{i_\nu i_{\nu+1}}$ is $q_{ij}$, then $i_\nu$ and
$i_{\nu+1}$ are even. Since equally many odd and even indices remain
after the factors from the diagonal submatrices were removed, the
product has equally many factors $p_{ij}$ and $q_{ij}$. I denote by
$m$ this number, which takes the values of the variable $m$ in
\eqref{eq:omega}, the integral values from 0 to $\lfloor d/2 \rfloor$.
The general term in \eqref{eq:pf} now is
\begin{multline}\label{eq:term}
  \sgn \begin{pmatrix} 2k_1-1 & 2k_1 & 2k_2-1 & \dots &2k_{2m}\\
    i_1 & i_2 & i_3 & \dots & i_{4m} \end{pmatrix} \\
    \prod_{\text{odd $\nu < 4m$}} s_{i_\nu i_{\nu+1}} ,
\end{multline}
where $1 \le k_1 < k_2 < \cdots < k_{2m} \le d$, and for every $\nu$,
the indices $i_\nu$ and $i_{\nu+1}$ are either both odd or both even.
For a given $m$, the sum runs over all such ordered sets
$(k_i \mid i = 1 \dots 2m)$ and all partitions of
$\{2k_1 - 1 , 2k_1 , \linebreak 2k_2 - 1 , \dots , 2k_{2m} \}$ into
such pairs $\{i_\nu,i_{\nu+1}\}$.

A sequence of cyclic permutations of odd length changes the sequence
in the upper row of the permutation symbol in \eqref{eq:term} into
$2k_1 - 1 , 2k_2 - 1 , \dots , 2k_{2m} - 1 , \linebreak 2k_{2m} ,
\dots , 2k_2 , 2k_1$. This leaves the signature unaltered. Also
without changing the signature, one can reorder the pairs
$\{i_\nu,i_{\nu+1}\}$ in the lower row so that all the pairs with odd
$i_\nu$ and $i_{\nu+1}$ appear before the pairs with even $i_\nu$ and
$i_{\nu+1}$. Flipping $i_\nu$ and $i_{\nu+1}$ in the latter pairs is
equivalent to replacing every $q_{ij}$ by $q_{ji}$. After these
permutations, one can write the entries in the first half of the lower
row of the permutation symbol in the form $2i_1-1,2i_2-1,\dots,2i_m-1$
and the entries in the second half of that row in the form
$2j_m,\dots,2j_2,2j_1$. The signature in \eqref{eq:term} thus becomes
\begin{multline}
  \sgn \begin{pmatrix}k_1&k_2&\cdots&k_{2m}\\i_1&i_2&\cdots&i_{2m}
  \end{pmatrix}
  \sgn \begin{pmatrix}k_1&k_2&\cdots&k_{2m}\\j_1&j_2&\cdots&j_{2m}
  \end{pmatrix} \\
  = \sgn \begin{pmatrix}i_1&i_2&\cdots&i_{2m}\\j_1&j_2&\cdots&j_{2m}
    \end{pmatrix} ,
\end{multline}
and the product becomes 
\begin{equation}
  \prod_{\text{odd $\nu < 2m$}}  p_{i_\nu i_{\nu+1}} q_{j_{\nu+1
      j_\nu}} .
\end{equation}
Since both sets
$\{\{i_1,i_2\},\{i_3,i_4\},\dots,\{i_{2m-1},i_{2m}\}\}$ and
$\{\{j_1,j_2\}\{j_3,j_4\},\dots,\{j_{2m-1},j_{2m}\}\}$ take the values
of all partitions of a common even subset of $\{1,2,\dots,d\}$ into
pairs, by comparison with \eqref{eq:omega}, one arrives at
\eqref{eq:S}.

\section{\label{sec:sum}Summary}

The representation of a Bogolyubov transformation of fermion
annihilation and creation operators on the Fock space related to a
finite-dimensional space of single-fermion states was discussed from
the point of view of its equivalence to a spin representation of an
orthogonal group. It was shown, in particular, that a much discussed
``sign problem of the Onishi formula'' can be traced back to the
double-valuedness of spin representations. The sign ambiguity referred
to by this language affects a formula for the overlap amplitude
between two quasi-fermion vacua in the much cited book by Ring and
Schuck and some related formulas but not the original formula of
Onishi and Yoshida, whose is scope is, however, more limited.
Derivations based on the interpretation of the Fock space
representation as a spin representation were show for the formula of
Ring and Schuck and some related formulas. In some cases, these
derivations are more complete than those in the literature. I gave a
short proof of the formula of Onishi and Yoshida, whose derivation is
missing in their paper, based on a relation written there, and a new,
combinatoric, proof of an equivalent formula recently derived by
Robledo.

\bibliography{bogo}

\end{document}